\documentclass[prd,nofootinbib,aps,superscriptaddress,tightenlines,preprintnumbers,twocolumn]{revtex4}

\usepackage{epsfig,latexsym,amsmath,amssymb}
\usepackage{hyperref}
\def\a{\alpha}		\def\b{\beta}		
\def\d{\delta}		\def\e{\epsilon}		
\def\g{\gamma}				
			\def\k{\kappa}		\def\l{\lambda}
\def\m{\mu}		\def\n{\nu}			\def\o{\omega}
					\def\r{\rho}

\def\L{\Lambda}

\def\uf{v} % full fluid velocity

\def\lag{{\mathcal{L}}}

\def\ki{{\bar{k}_i}}
\def\kj{{\bar{k}_j}}
\def\k3{{\bar{k}_3}}
\def\kpb{{\bar{k}_{\bot}}}

\def\dr{\delta_{\rho}} % density contrast
\def\vf{U} %vector field perturbation
\def\duf{V} %fluid velocity perturbation
\begin{document}

\preprint{\hbox{CALT-68-2683}} 

\title{Direction-dependent Jeans instability in an anisotropic Bianchi type I space-time}
\author{Timothy R. Dulaney}
\email[]{dulaney@theory.caltech.edu}
\affiliation{California Institute of Technology, Pasadena, CA 91125}

\author{Moira I. Gresham}
\email[]{moira@theory.caltech.edu}
\affiliation{California Institute of Technology, Pasadena, CA 91125}

\begin{abstract}
We derive the metric for a Bianchi type I space-time with energy density that is dominated by that of a perfect fluid with equation of state $p=w\rho$ and whose anisotropy is seeded by a fixed norm spacelike vector field.   We solve for the evolution of perturbations about this space-time.  In particular, the Jeans instability in an expanding flat Friedmann-Robertson-Walker universe is modified by the presence of the vector field so that energy density perturbations develop direction-dependent growth.  We also briefly consider observational limits on the vector field vacuum expectation value, $m$.    We find that, if $m$ is constant during recombination and thereafter, $m \lesssim 10^{14} GeV$.
\end{abstract}

\date\today
\maketitle

%%%%%%%%%%%%%%%%%%%%%%%%%%%%%%%%%%%%%%%%%
\section{Introduction}

The Universe is usually assumed to be homogeneous and isotropic on large scales. 
In this paper, we make progress in testing the assumption of isotropy by exploring a particular model of isotropy-breaking involving fixed-norm spacelike vector fields minimally coupled to gravity.
Asymmetries (like anisotropy) that have been very nearly ruled out in the current cosmological epoch could have been present in the early Universe; the signatures of asymmetries long ago could be very subtle and therefore not ruled out experimentally. Furthermore, although rotational symmetry breaking in Standard Model particle interactions is very tightly constrained (see, for example, Refs.~\cite{Kostelecky:1999mr, Bluhm:1999dx,  Bluhm:1999ev, Kostelecky:2001mb, Kostelecky:2002hh, Cane:2003wp, Bailey:2004na}), isotropy-breaking through minimal coupling to gravity is not as well-constrained 
\cite{Moore:2001bv, Burgess:2002tb, Kostelecky:2003fs, Bailey:2006fd}.

In trying to understand a broken symmetry, it is often convenient to study a particular model that provides the mechanism for symmetry breaking. There is a class of natural models for breaking Lorentz invariance spontaneously involving fixed-norm vector fields. 
Fixed-norm timelike vector field models that break Lorentz invariance but not rotational invariance have been extensively studied \cite{Kostelecky:1989jw, Jacobson:2004ts, Carroll:2004ai, Lim:2004js, Eling:2003rd, Foster:2005dk, Li:2007vz, Seifert:2007fr, Foster:2006az, Eling:2006df, Jacobson:2000xp, Clayton:2001vy, Graesser:2005bg, Kanno:2006ty, Bluhm:2007bd}. For a review, see \cite{Jacobson:2008aj}. 

Recently, there has been interest in the consequences of a small breaking of rotational invariance during the inflationary era \cite {Barrow:1997sy, Ackerman:2007, Gumrukcuoglu:2007bx, Gumrukcuoglu:2006xj, Pullen:2007tu, Ando:2007hc, Boehmer:2007ut, Pitrou:2008gk}.  
A variant of the models of Refs.~\cite{Carroll:2004ai}, \cite{Jacobson:2000xp}, and \cite{Kostelecky:1989jw}  provide a convenient framework with which to explore such consequences \cite{Ackerman:2007, Dulaney:2008ph}. The model involves a spacelike, fixed-norm vector field, $u^\m$ with Lagrange density~\cite{Carroll:2004ai},
\begin{align}
\label{ulag}
  {\cal{L}}_u 	&= -\beta_1 \nabla^\mu u^\sigma \nabla_\mu u_\sigma - \beta_2 (\nabla_\mu u^\mu)^2 \nonumber \\  
  			&- \beta_3 \nabla^\mu u^\sigma \nabla_\sigma u_\mu + \lambda(u^{\mu} u_{\mu} - m^2)\ .
\end{align}
Here $\lambda$ is a Lagrange multiplier that enforces the fixed norm constraint $ g_{\mu \nu} u^{\mu} u^{\nu}=m^2$.   The Lagrange multiplier term can be thought of as encoding the energy-minimizing effects of a smooth potential.  The four-vector $u^{\sigma}$ is spacelike and induces the spontaneous breaking of rotational invariance when $m^2>0$. (A timelike four-vector would break Lorentz invariance but not necessarily rotational invariance.) A model for the breaking of isotropy during the inflationary era with dynamics governed by the action,
\begin{equation}
\label{dyn}
S=\int {\rm d}^4x {\sqrt{-g}}\left({1 \over 16 \pi G} R - \rho_{\L}  +  {\cal{L}}_u\right),
\end{equation}
where $\rho_\L$ is a constant energy density, was considered in \cite{Ackerman:2007} in the interest of understanding effects of isotropy breaking during inflation on the cosmic microwave background (CMB).\footnote{A related $5$-dimensional model was recently considered in \cite{Carroll:2008pk} as a way to hide large extra dimensions.} When coordinates are chosen such that the $x_3$-axis is parallel to the direction along which the vector field gets its vacuum expectation value (VEV), the resultant space-time is given by \cite{Ackerman:2007},
\begin{equation}
\label{coord}
{\rm d}s^2=-{\rm d}t^2+a(t)^2{\rm d}{\bf x}_{\perp}^2+b(t)^2{\rm d}x_3^2,
\end{equation}
where,
\begin{equation}
a(t)=e^{H_a t},~~~b(t)=e^{H_b t},
\label{spit1}
\end{equation} 
and,
\begin{align}
H_a &= {{\dot a}\over a} = H_b(1+16 \pi G  \beta_1 m^2),\nonumber  \\
H_b & = {{\dot b}\over b} ={{\sqrt{ 8 \pi G \rho_{\Lambda} \over (1+8 \pi G  \beta_1 m^2)(3+32
\pi G  \beta_1 m^2)}}} \,. \label{backgrounda}
\end{align}

In the quest toward testing the standard assumption of spatial isotropy in our Universe, it is informative to consider the existence and properties of other anisotropic space-times seeded by a fixed-norm vector field. To this end, we consider a universe with a fixed-norm spacelike vector field, permeated by a perfect fluid with equation of state $p = w \rho$ where $w$ is constant.   We consider this perfect fluid case because (1) it can model relevant cosmological epochs such as radiation ($w = {1\over 3}$) and cold non-relativistic matter ($w = 0$) domination and (2) it is the next most mathematically tractable example of an anisotropic space-time.  The case $ w = -1$ corresponds to a vacuum energy density as considered in~\cite{Dulaney:2008ph}. 

We assume that the space-time geometry is approximately spatially flat; Ref.~\cite{Hao:2003jm} found that an approximately spatially flat geometry is preferred if anisotropy is to remain small. 
Considering this particular model of spontaneous isotropy-breaking when ordinary matter is present allows us to study the modified behavior of matter and metric perturbations---in particular the modification of the Jeans instability. This particular model also provides an example of an expanding Bianchi type I space-time that does not evolve to be isotropic. Also, the deviation from isotropy can be tuned by a single parameter.

In Ref.~\cite{Dulaney:2008ph}, an analysis of vector field perturbations in flat space led to the requirement for energetic stability: $\b_1 + \b_2 + \b_3 = 0$ and $\b_1 > 0$. This result carries over to our present analysis (since the analysis in \cite{Dulaney:2008ph} was independent of gravity and other matter fields). We shall further suppose that an analysis of energies in non-Minkowski space would lead to the requirement that the vector field Lagrange density take the form of a field strength tensor squared ($ \lag_u \sim F_{\m \n} F^{\m \n} $) in both Minkowski and non-Minkowski space-time.\footnote{In Minkowski space, $\b_1 + \b_2 + \b_3 = 0$ corresponds to a field strength squared Lagrange density, while in curved space a field strength squared Lagrange density corresponds to the more restricted case, $\b_1 = -\b_3$ and $\b_2 = 0$.} Therefore, we consider,
\begin{equation}
\label{vector field lagrange density}
  {\cal{L}}_u = -{1 \over 4} F^{\m\n} F_{\m\n}  + \lambda(u^{\mu} u_{\mu} - m^2),
\end{equation}
where $F_{\m\n} = \nabla_\m u_\n - \nabla_\n u_\m$.  We have set $\beta_{1}=1/2$ in order to canonically normalize the kinetic term.  Thus the parameter $m^{2}$ completely characterizes the deviation from isotropy.

%%%%%%%%%%%%%%%%%%%%%%%%%%%%%%%%%%%%%%%%%

\section{Background space-time}\label{background}

We take the following action to govern our model,
\begin{equation}
\label{dyn}
S=\int {\rm d}^4x {\sqrt{-g}}\left({1 \over 16 \pi G} R + {\cal{L}}_{fluid} +  {\cal{L}}_u\right),
\end{equation}
where ${\cal L}_u $ is given by Eq.~\eqref {vector field lagrange density}. 

The energy-momentum tensor for $u^\mu$ derived from (\ref{vector field lagrange density}) is \cite{Carroll:2004ai},
\begin{align} 
T_{\m\n}^{(u)} &= F_{\m\rho}F_{\n}{}^{\rho} +g_{\mu\nu}{\cal{L}}_u+u_{\m}\left(g_{\n\sigma} -{u_{\n} u_{\sigma} \over m^{2} }\right)\nabla_{\rho} F^{\rho\sigma} \nonumber \\
&+ u_{\n}\left(g_{\m\sigma} -{u_{\m} u_{\sigma} \over m^{2} }\right)\nabla_{\rho} F^{\rho\sigma} + {u_{\m}u_{\n} \over m^2}u_{\sigma}\nabla_{\rho} F^{\rho\sigma}.  \label{tmunu}
\end{align}

The fluid stress energy tensor is given by,
\begin{equation}
T_{\mu\nu}^{(fl)} = (\rho + p)\uf_\m \uf_\n + p g_{\m \n},
\end{equation}
where $\uf_\m$ is the fluid's four-velocity and $\rho$ and $p$ are the energy density and pressure in the fluid's rest frame.  Taking the equation of state to be of the form $p = w \rho$, the stress-energy tensor takes the form,
\begin{equation}
T_{\mu\nu}^{(fl)} = \rho \left[ (1+w) \uf_\m \uf_\n + w g_{\m \n} \right].
\end{equation}
We assume the fluid is homogeneous and isotropic in its rest frame, so $\rho = \rho(t)$.  

\subsection{Solving Einstein's equations}
We choose coordinates such that the fluid is at rest. Since the fluid is isotropic, we still have the freedom to choose spatial coordinates such that the $x_3$ axis aligns with the isotropy-breaking vector field . However, a priori, the vector field may still have a timelike component in the rest frame of the fluid.  But, as is natural, we will assume that the vector field aligns such that it has no timelike component in the rest frame of the fluid.  We hope to address this alignment issue in a future publication.  

Taking the vector field to have no timelike component, we have,
\begin{equation} \label{uvector}
\bar{u}_\m = m \sqrt{\bar{g}_{33}} \eta_{\m 3},
\end{equation} 
where $\bar{g}_{\m \n}$ is the background metric.  We make the ansatz for the background metric,
\begin{equation}
ds^2 = -dt^2 + a^2(t) d \mathbf{x}_\perp^2 + b^2(t) (dx^3)^2, \label{bkgnd metric}
\end{equation}
where $d \mathbf{x}_\perp^2 = (dx^1)^2 + (dx^2)^2$.  This is an axisymmetric Bianchi type I geometry.  Given Eqs.~\eqref{uvector} and \eqref{tmunu}, the nonvanishing components of the background stress tensor for the vector field are \cite{Ackerman:2007},
\begin{eqnarray}
\bar{T}_{00}^{(u)} &=&  {1 \over 2} m^2\left({\dot{b}\over b}\right)^2,
\nonumber\\
\bar{T}_{11}^{(u)}&=& \bar{T}_{22}^{(u)}=  {1 \over 2} m^2 a^2 \left({\dot{b}\over b}\right)^2,
\\
\bar{T}_{33}^{(u)} &=&   {1 \over 2} m^2 b^2 \left(  \left( {\dot{b} \over b} \right)^2 - 2 \left( {\ddot{b} \over b} \right)  -4 \left( {\dot{a} \over a } \right) \left( {\dot{b} \over b}\right) \right). \nonumber
\end{eqnarray}
Note that the background solution satisfies the weak energy condition if,
\begin{equation}
1+w \ge {4 \epsilon \over 3 + 4\epsilon},
\end{equation}
where we have defined,
\begin{equation}
\epsilon \equiv 4 \pi G m^2.
\end{equation}

The field equations,
\begin{equation}
G_{\mu \nu} = 8 \pi G \left(T^{(u)}_{\m \n} + T^{(fl)}_{\m \n}\right),
\end{equation}
give the following non-trivial equations ($\m = \n = 0$, $\m = \n = 1$ and $\m = \n =3$ respectively) for the background:
\begin{align}
&\left({\dot{a} \over a}\right)^2 + 2 \left({\dot{a} \over a}\right) \left({\dot{b} \over b}\right)= 8\pi G \bar{\rho}(t) + \epsilon \left({\dot{b} \over b}\right)^2, \label{bfe1}\\
&\left({\ddot{a} \over a}\right)+\left({\ddot{b} \over b}\right) +\left({\dot{a} \over a}\right)\left({\dot{b} \over b}\right) = -8\pi G w \bar{\rho}(t) - \epsilon \left({\dot{b} \over b}\right)^2, \label{bfe2} \\
&2\left({\ddot{a} \over a}\right) + \left({\dot{a} \over a}\right)^2 = -8 \pi G w \bar{\rho}(t) \nonumber \\ 
&- \epsilon \left[\left({\dot{b} \over b}\right)^2 - 2 \left({\ddot{b} \over b}\right) -4\left({\dot{a} \over a}\right)\left({\dot{b} \over b}\right) \right].\label{bfe3}
\end{align}
Although it is not independent, it is helpful to consider the continuity equation,
\begin{align}
0 &= \nabla^\m  \left(T^{(u)}_{\m 0} + T^{(fl)}_{\m 0}\right)\vert_{\text{background}} \nonumber \\ 
&= \dot{\bar{\rho}} + (1+w) \bar{\rho} \left(2 \left({\dot{a} \over a}\right) + \left({\dot{b} \over b}\right) \right).
\end{align}
We can simplify this equation in the usual way to determine the relationship between the scale factors and the energy density $\bar{\rho}$,
\begin{align}
{\dot{\bar{\rho}} \over \bar{\rho}} &= -(1+w) \left(2 \left({\dot{a} \over a}\right) + \left({\dot{b} \over b}\right) \right) \nonumber \\
&= -(1+w) {1 \over a^2 b} {d \over dt} \left(a^2 b\right).
\end{align}
The solution for the time dependence of $\bar{\rho}$ is given by,
\begin{equation} \label{rho}
{\bar{\rho}(t) \over \bar{\rho}(t_0)} = \left( { a^2(t_0) b(t_0) \over a^2(t) b(t)} \right)^{1+w}.
\end{equation}  
 
There are two linear combinations of the field equations \eqref{bfe1}-\eqref{bfe3} that are independent of $\bar{\rho}$:
\begin{align}
\left({\ddot{a} \over a}\right)+ \left({\ddot{b} \over b}\right) &=  \epsilon (w -1) \left({\dot{b} \over b}\right)^2~\nonumber \\
 &-(2w+1) \left({\dot{a} \over a}\right) \left({\dot{b} \over b}\right) - w \left({\dot{a} \over a}\right)^2,  \\
\left({\ddot{a} \over a}\right) - \left({\ddot{b} \over b}\right) &= 2 \epsilon \left[\left({\ddot{b} \over b}\right) +2\left({\dot{a} \over a}\right)\left({\dot{b} \over b}\right) \right] \nonumber \\
&- \left({\dot{a} \over a}\right)\left(\left({\dot{a} \over a}\right)  - \left({\dot{b} \over b}\right) \right) .
\end{align}
It is clear from the form of these linear combinations that the solution is of the form $a(t) \propto t^\a$ and $b(t) \propto t^\b$. A third independent equation implies $\bar{\rho}(t) \propto t^{-2}$  (when $w \neq -1$) and thus, by Eq.~\eqref{rho},
\begin{equation}
(2 \a + \b)(1 + w) = 2.
\end{equation}
We find from these equations that,
\begin{align}
\a &= {2(1+2\epsilon) \over (1+w)(3+4\epsilon)}, \\
\b &= {2 \over (1+w)(3+4\epsilon)}.
\end{align} 

The background metric satisfies the following equations,
\begin{align}
H_a(t) &= {{\dot a}\over a} = H_b(t)(1+2 \epsilon ) \label{Ha},   \\
H_b(t) &= {{\dot b}\over b} = \sqrt{{8\pi G \bar{\rho}(t) \over (1 +  \epsilon)(3 + 4 \epsilon)}} \label{Lambda}.
\end{align}
These relationships between $H_a$, $H_b$, and $\bar{\rho}$ are identical in form to the relationships, Eqs.~\eqref{backgrounda}, that were found in Ref.~\cite{Ackerman:2007} for the case of a deSitter-like space-time (\emph{i.e.}~$w = -1$).  For $w \neq -1$, we have just found that 
\begin{equation} \label{bkgd rho}
\bar{\rho}(t) = \bar{\rho}(t_0) \left({t_0 \over t}\right)^2,
\end{equation}
as in the case of a flat FRW universe.\footnote{This indicates that evolution equations parameterized by cosmic time do not have a smooth limit $w \rightarrow -1$, since for $w=-1$ the energy density is constant.}  However, since $\epsilon > 0$, the direction in which the fixed-norm spacelike vector gets a VEV  expands more slowly than in the isotropic case, while the transverse directions expand more quickly. We assume the deviation from isotropy will be small, so $\epsilon <<1$; in this limit,
\begin{align}
{\dot{a} \over a} &\simeq {{2 \over 3(1 + w)t}\left(1 + {2 \over 3} \epsilon \right)}, \\
{\dot{b} \over b} &\simeq {{2 \over 3(1 + w) t}\left(1 - {4 \over 3} \epsilon  \right)}.
\end{align} 
In the isotropic limit ($\epsilon \rightarrow 0$), the flat FRW metric for a perfect fluid with equation of state $p = w \rho$ is recovered.  

%%%%%%%%%%%%%%%%%%%%%%%%%%%%%%%%%%%%%%%%%

\section{Classical stability}\label{stability}
A viable model of isotropy breaking must be at least classically stable against small perturbations. In~\cite{Dulaney:2008ph}, it was shown that the space-time given by Eqs.~\eqref{coord}-\eqref{backgrounda} is classically stable against small perturbations for a range of parameters. More specifically, the authors analyzed the equations of motion for vector field perturbations in flat space and showed that when the parameters in Eq.~\eqref{ulag} satisfy $\b_1 + \b_2 + \b_3 = 0$ and $\b_1 > 0$, the energy of field configurations that satisfy the Euler-Lagrange equations of motion is non-negative at all orders in perturbations.
When $\b_1 + \b_2 + \b_3 \neq 0$, negative energy modes propagate. Thus models with $\b_1 + \b_2 + \b_3 \neq 0$ were deemed unstable. Also, Einstein's equations were solved in the limit where mode wavelengths are much smaller than the Hubble radius, and in the opposite limit where mode wavelengths are much longer than the Hubble radius. Modes in the short wavelength limit were shown to propagate as luminal plane waves with constant amplitudes. Short wavelength modes with superluminal phase velocities were found to propagate unless $\b_1 + \b_3 = 0$ and $\b_2 = 0$. Modes in the long wavelength limit were shown to have only constant parts and parts that decay as $e^{-(2 H_a + H_b) t} = 1 / (a^2 (t) b(t))$ or $e^{- 16 \pi G m^2 \b_1 H_b t} = b(t) / a(t) $. 

We now consider small perturbations about the background given by Eqs.~\eqref{uvector}, \eqref{bkgnd metric}, \eqref{Ha}, and \eqref{Lambda}. 
However, let us first recall the story of perturbations in an expanding, flat FRW universe. For perfect fluids with constant equation of state, physical (\emph{i.e.}~gauge invariant) perturbations can behave differently depending on the scale of the wavelength of the perturbation.\footnote{Homogeneity of the space-time allows for perturbations to be Fourier-transformed in space and guarantees that modes with different Fourier wave numbers will evolve independently.} Modes with wavelengths shorter than the Jeans wavelength (which is on the order of $\sqrt{w} H^{-1}$) will oscillate as plane waves. Modes with wavelengths much larger than $H^{-1}$ decay or remain constant. Modes---in particular, the energy density perturbation---at the intermediate scale can grow or decay; this is the Jeans instability in an expanding universe. The instability is most efficient when $w \approx 0$ since the Jeans wavelength goes to zero when $w \rightarrow 0$.

We study only the very short wavelength (\emph{i.e.}~$ |\sqrt{w} k| \gg H$) and long wavelength (\emph{i.e.}~$|\sqrt{w} k| \ll H$) limits. However, when $w = 0$, the very short wavelength limit as stated does not exist. For $w = 0$, we study the limit, $k \gg H$; the Jeans instability is manifest at this scale. Considering the very short wavelength and long wavelength limits is enough to establish stability as compared to the isotropic case. We assume that we can glean the interesting physics of the Jeans instability in an anisotropic FRW-like universe by considering an intermediate (short) wavelength scale in only the $w = 0$ case.

\subsection{Parameterizing independent perturbations}

We define the ten (comoving) metric perturbations by ($h_{\m\n}=h_{\n\m}$), 
\begin{align}
ds^2 &= - (1 + h_{00})\, dt^2 + 2 a(t) \,h_{0 i} \,dt dx^i  \nonumber \\
&+ 2 b(t)\, h_{0 3} \, dt dx^3 + 2 a(t) \,b(t) \,h_{i 3} dx^i dx^3  \label{perturbed metric}\\
&+ a(t)^2 \, (\delta_{i j} + h_{i j}) dx^i dx^j + b(t)^{2}(1+ h_{33}) (dx^3)^{2}. \nonumber
\end{align}
We define the vector field perturbations and fluid velocity perturbations by,
\begin{align} \label{perturbed u}
 \d u^\m &= u^\m - \bar{u}^\m, \\  \label{perturbed v}
 \d \uf^\m &= \uf^\m - \bar{\uf}^\m, \qquad \bar{\uf}^\m = - \bar{g}^{0 \m}.
\end{align}

We make the approximation that the fluid's equation of state is unperturbed (so $ p - \bar{p} = w  \bar{\rho} \d_{\rho} $) and that the shear stress perturbations are negligible compared to the pressure perturbations.\footnote{Assuming $\d \rho = w \d p$ is equivalent to assuming that the perturbations are adiabatic. See, for example Chapter 6 in Ref.~\cite{Mukhanov}.}   This is an excellent approximation for a matter dominated era and a radiation dominated era before neutrino decoupling. (See, for example,~\cite{Ma:1995ey, Bertschinger:2001is}.)

Since $\uf^\m \uf_\m = -1$, the timelike fluid velocity perturbation is given by, $\d \uf^0 = -(1/2) h_{00} $, leaving only three independent fluid velocity field components.  Also, the vector field constraint, $u_\m u^\m = m^2$ implies that $\d u^3 = -(m/2b) h_{33}$, so there are only three independent vector field perturbations. 

It is convenient to define the comoving fluid velocity perturbations (\emph{i.e.}~the peculiar velocity of the fluid) by ($ i \in \{ 1, 2 \})$:
\begin{equation}
 \d \uf^i = {\duf^i \over a(t)}, \qquad \d \uf^3 = {\duf^3 \over b(t)}.
\end{equation}
Similarly we define the comoving vector field perturbations,
\begin{equation}
\delta u^{0} = m \vf^{0}, \qquad \delta u^{i} = m {\vf^{i} \over a(t)}. 
\end{equation}
Finally, we define the fluid energy density contrast by,
\begin{equation}
 \d_{\rho} = {\rho \over \bar{\rho} }- 1.
\end{equation}

%%%%%%%%%%%%%%%%%%%%%%%%%%%%%%%%%%%%%%%%%

\subsection{Gauge choice} \label{Gauge Choice}

We can use diffeomorphism invariance to remove four of the seventeen perturbations just discussed.  Under an infinitesimal coordinate transformation, $x^\mu \rightarrow x^\m + \xi^\m$, the vector field transforms to $u^{\lambda}+\Delta u^{\lambda}$ where,
\begin{equation} \label{uGauge}
\Delta u^{\lambda} = \bar{u}^{\mu}\partial_{\mu} \xi^{\lambda} -\xi^{\mu}\partial_{\mu}\bar{u}^{\lambda},
\end{equation}
and similarly $\uf^\m \rightarrow \uf^\m + \Delta \uf^\m$ where,
\begin{equation} \label{vGauge}
\Delta \uf^{\lambda}= \bar{\uf}^{\mu}\partial_{\mu} \xi^{\lambda} -\xi^{\mu}\partial_{\mu} \bar{\uf}^{\lambda}.
\end{equation}
The metric transforms to $g_{\mu \nu}+\Delta g_{\mu \nu}$ where,
\begin{equation} \label{metricGauge}
\Delta g_{\mu \nu}= -(\bar{\nabla}_\m \xi_\n + \bar{\nabla}_\n \xi_\m).
\end{equation}

In analyzing short wavlength perturbations, we choose a gauge where  the first order fluctuations in the (contravariant) four-vector field about their background, $\bar{u}^\m =  \eta^{\m 3} m/b(t)$, vanish (\textit{i.e.}~$\d u^\mu = 0$). This gauge condition is satisfied by fixing, 
\begin{align}
- \d(u^\l) &= \Delta u^\l \vert_{1^{st} \text{order}}  \nonumber	\\
	&= \bar{u}^\m \partial_\m \xi^\lambda - \xi^\m \partial_\m \bar{u}^\l \nonumber  \\
	&= {m \over b(t)} \partial_3 \xi^\l + \delta^{\l}_{3} \xi^0 {m \over b(t)} H_b.
\end{align} 
We shall refer to this as \textit{vector field gauge}.  Using this gauge allows us to carry over some results from \cite{Dulaney:2008ph} since the vector field gauge was also used in that analysis.  

We find it more convenient to use a \textit{modified synchronous gauge},
\begin{equation}
h_{0i} = h_{03} =0 \qquad \text{and} \qquad \vf^{3} = 0,
\end{equation} 
in the analysis of long wavelength perturbations. This gauge condition is satisfied by fixing, 
\begin{align}
- \d(u^3) 	&= \Delta u^\l = {m \over b(t)} \partial_3 \xi^3 +  \xi^0 {m \over b(t)} H_b,
\end{align} 
\begin{align}
- \d(g_{0 i}) &= - a(t) h_{0 i} =  \Delta g_{0 i}  \nonumber \\
&= - \partial_0 \xi_i - \partial_i \xi_0 + 2 H_a \xi_i,
\end{align}
and,
\begin{align}
- \d(g_{0 3}) &= - b(t) h_{0 3}=  \Delta g_{0 3}  \nonumber \\
&= - \partial_0 \xi_3 - \partial_3 \xi_0 + 2 H_b \xi_3.
\end{align}

The vector field satisfies the Lagrange multiplier equation, $u^2 = m^2$.  This equation and either gauge choice imply that,
\begin{equation}
(\d g_{\m\n}) \bar{u}^\m \bar{u}^\n + 2 \bar{g}_{\m\n} (\d u^\m) \bar{u}^\n  = h_{33}m^2  = 0.
\end{equation}

%%%%%%%%%%%%%%%%%%%%%%%%%%%%%%%%%%%%%%%%%

\subsection{Perturbation equations}
By the twice-contracted Bianchi identity, conservation of total energy-momentum density follows from Einstein's equations. 
Also, one can check that the equations of motion for the vector field are equivalent to the conservation of the vector field stress-energy tensor. In general, $\nabla_\m {{T^{(u)}}^\m}_3 = 0$ identically at first order in perturbations, while the other three vector field stress-energy conservation equations are nontrivial. This---along with the twice-contracted Bianchi identity---means that there are only thirteen independent equations.  The first order contracted Bianchi identities can be written as,
\begin{equation}
8 \pi G \delta(\nabla_{\mu}( {{T^{(fl)}}^\m}_{\nu}+ {{T^{(u)}}^\m}_{\nu})) =  \nabla^{\mu} \delta E_{\mu \nu},
\end{equation}
where,
\begin{equation}
\delta E_{\mu \nu} \equiv  \delta (G_{\mu \nu}) - 8 \pi G \delta (T^{(fl)}_{\mu \nu} + T^{(u)}_{\mu \nu} ). 
\end{equation}

See Appendix A for the full set of equations without any gauge condition applied.   

%%%%%%%%%%%%%%%%%%%%%%%%%%%%%%%%%%%%%%%%%

\subsection{Short wavelength fluid dynamics and Jeans instability}

Einstein's equations in the very short wavelength limit ($\sqrt{w} k \gg a H$) are independent of $\bar{\rho}$ and the background metric. Thus, since we use the same gauge as in \cite {Dulaney:2008ph}, the solutions for metric perturbations in the short wavelength limit found there (and described at the beginning of this section) apply also to the present case of a perfect fluid. The fluid pertubations must have the same plane-wave form as the metric perturbations when the metric perturbations are non-zero. The fluid stress-energy conservation equations relate the metric perturbation amplitudes to the fluid perturbation amplitudes. 

In principle, there can also be a non-gravitational fluid perturbation mode governed by the fluid equations in flat space.  The first order perturbed fluid stress-energy conservation equations (in flat space) are:\footnote{For a review of hydrodynamics in the absence of gravity see, for example,~\cite{MTW:1973}.}
\begin{align}
0 &=(1+w) \vec{\nabla} \cdot \vec{\duf} + \partial_t \d_\rho, \label{fluid 0}\\
0 &=w \vec{\nabla} \d_\rho + (1+w) \partial_t \vec{\duf}. \label{fluid i}
\end{align}
We may decompose $\vec{\duf}$ as:
$$
\vec{\duf} = \vec{\nabla} \Phi + \vec{B} \; ; \; \vec{\nabla} \cdot \vec{B} = 0.
$$

Then taking the time derivative of Eq.~\eqref{fluid 0} and the divergence of Eq.~\eqref{fluid i} and substituting in for $\nabla \cdot \vec{\duf}$ gives,
\begin{equation}
(w \nabla^2 - \partial_t^2) \d_\rho=0,
\end{equation}
while taking the gradient of Eq.~\eqref{fluid 0}  and the time derivative of Eq.~\eqref{fluid i} and similarly substituting gives,
\begin{equation}
(w \nabla^2 - \partial_t^2) \vec{\nabla} \Phi - \partial_t^2 \vec{B} = 0.
\end{equation}

The solutions for $\Phi$ and $\d_\rho$ are plane waves,
\begin{equation}\label{sw rho}
\d_\rho = A_\rho e^{i \o t \pm i \vec{k} \cdot \vec{x}} \; \text{and} \; \vec{\nabla}\Phi = \vec{A}_\Phi e^{i \o t \pm i \vec{k} \cdot \vec{x}},
\end{equation}
where, 
\begin{equation}\label{sw omega}
 \o^2 = w ( \vec{k}\cdot \vec{k} ) \;\; \text{and} \;  \vec{A}_\Phi = \mp {\sqrt{w} \over 1 + w} {\vec{k} \over | \vec{k} |} A_\rho.
\end{equation}
The divergence-free part of the fluid velocity, $\vec{B}$, must be independent of time.  We recover the standard result: the speed of sound of the perfect fluid is $\sqrt{w}$.

In the case of a cosmological constant ($w=-1$) the energy density perturbation is independent of space-time coordinates and thus remains a constant.  

Special attention must be paid to the case $w = 0$.  It was assumed that $a \partial_{t} \sim k$ in the short wavelength analysis of \cite{Dulaney:2008ph}, but  Eqs.~\eqref{sw rho} and \eqref{sw omega} show that this is not a valid assumption for modes with non-zero energy density contrast. If $\partial_{t} \sim H_{a,b}$, then none of the ten Einstein's equations, $\d E_{\m \n}$, nor the fluid energy-density conservation equations (see Appendix A) constrain $\d_{\r}$ in the limit $k \gg H$ when $w = 0$. To find the evolution of $\d_\r$, we must find a $\bar{k}$-independent linear combination of the equations that involves $\d_\r$.  We have found such equations in the cases $k_{3} = 0$ and $k_{\perp} = 0$.

When $k_{\perp} \neq 0$ and $k_{3} =0$, one finds the following $\bar{k}$-independent equation, 
\begin{multline}
\left(-{1\over2}(3+4\epsilon)H_{b}^{2} + 2H_{b}(1+2\epsilon)\partial_{t} + \partial_{t}^{2} \right)\d_{\rho}  \\
- \left({1\over2}(3+4\epsilon)H_{b}^{2} -{1\over2}(3+4\epsilon)H_{b}\partial_{t}\right)h_{00}= 0.
\end{multline}
In the short wavelength limit, each of equations~\eqref{eij} and \eqref{fluid ti} tells us that $h_{00}$ vanishes; thus for this mode,
\begin{equation} \label{k3Jeans}
\ddot{\d}_{\rho} + 2H_{a}\dot{\d}_{\rho}= \left({4\pi G \over (1+\epsilon)}\bar{\rho}\right) \d_{\rho}.
\end{equation}
The solution of this equation is,
$$
\d_{\rho} = C_1 t^{2/(3+4\epsilon)} + C_2 t^{-1} .
$$ Note that  $t^{2/(3+4\epsilon)} \propto b(t)$ and $t^{-1} \propto H_{a,b}$.  

When $k_{\perp} = 0$ and $k_{3} \neq 0$, we find the $k$-independent equation,
\begin{multline}
\left( -(3 + 4 \epsilon)(1+\epsilon)H_b^{2} + 4 H_b \partial_t + 2 \partial_t^{2} \right) \d_\r \\
+ \left( -(3 + 10 \epsilon + 4 \epsilon^{2})H_b^{2} + 2(1 + \epsilon) H_b \partial_t \right)h_{00} \\
+ \left( 3 \epsilon H_b^{2} + ({3 /2} - \epsilon)H_b \partial_t + {1 \over 2} \partial_t^{2} \right)h_{i i} = 0.
\end{multline} 
In the short wavelength limit when $k_{\perp} = 0$, Eq.~\eqref{fluid t3} implies that $h_{00}$ vanishes while Eq.~\eqref{e00} implies that $h_{ii}$ vanishes.  Therefore, for this mode the density contrast evolves as,
\begin{equation} \label{kpJeans}
\ddot{\d}_{\rho} + 2 H_{b} \dot{\d}_{\rho} = (4\pi G \bar{\rho}) \d_{\rho}.
\end{equation}
The solution of this equation is,
$$
\d_{\rho} = C_1 t^{\g_+} + C_2 t^{\g_-},
$$ where, 
\begin{align}
\g_{\pm} &= \frac{-1 +4\epsilon \pm 5 \sqrt{1+3\epsilon(1+ \epsilon)(4/5)^{2}}}{2(3+4\epsilon)}, \\
&\simeq \left({-1 \pm 5 \over 6}\right) -{2 \over 9}\left(-4 \pm {7 \over 5}\right)\epsilon.
\end{align}

In the isotropic limit, $\epsilon \rightarrow 0$, 
\begin{equation}
\ddot{\d}_{\rho} + 2 H \dot{\d}_{\rho} = (4\pi G \bar{\rho}) \d_{\rho},
\end{equation}
where $H = \lim_{\epsilon \rightarrow 0} H_{a,b}$ and thus the Jeans instability for a flat FRW space-time is recovered: $\d_{\rho} \simeq C_1 t^{2/3} + C_2 t^{-1}$. The above solutions tell us that, in the anisotropic case, density perturbations in directions transverse to the direction in which the vector field gets a VEV grow more slowly while those modes with wavenumber in the parallel direction grow more quickly than in the isotropic case.  

%%%%%%%%%%%%%%%%%%%%%%%%%%%%%%%%%%%%%%%%%

\subsection{Long wavlength limit}
Having completed an analysis of short wavelength perturbations, we now consider the long wavelength limit of the perturbation equations.

Recall that $H_b$ is a function of time.  In the modified synchronous gauge, we have the following field equations in the long wavelength limit,

\begin{align}
{\delta E_{00} \over 1+\epsilon} &= -(3+4\epsilon) H_{b}^{2}(\delta_{\rho}+h_{00})+H_{b}\partial_{t} h_{jj} =0, 
\end{align}
\begin{align}
{\delta E_{0i} \over a(t)} &= (1+w)(3+4\epsilon)(1+\epsilon)H_{b}^{2}\duf^{i} =0, 
\end{align}
\begin{align}
{\delta E_{03} \over b(t)} &= (1+w)(3+4\epsilon)(1+\epsilon)H_{b}^{2}\duf^{3} =0,  
\end{align}
\begin{align} 
&{\delta E_{ij} \over a(t)^{2}} =\d_{ij} H_{b}(1+\epsilon)[-w(3+4\epsilon)H_{b}(\dr + h_{00})+\partial_{t}h_{00}]  \nonumber \\
& + {1 \over 2} \e_{l ( i} \e_{j ) k} \left( (3 + 4 \epsilon) H_b \partial_t + \partial_t^2 \right)h_{l k} = 0,
\end{align}
\begin{align}
&{2\delta E_{i3} \over a(t)b(t)} =\epsilon ( 3 - w(3 + 4 \epsilon) )H_{b}^{2}[(1+4\epsilon)h_{i3}+(2+4\epsilon)\vf^{i}] \nonumber \\
&+(1+4\epsilon)\left[ (3 + 4 \epsilon) H_{b}\partial_{t} + \partial_{t}^{2} \right] h_{i3} \\
&+4\epsilon[4(1 + \epsilon) H_{b} \partial_{t} + \partial_{t}^{2}] \vf^{i} = 0,  \nonumber
\end{align}
\begin{align}
&{\delta E_{33} \over b(t)^{2}} =H_{b}(1+\epsilon)[-w(3+4\epsilon)H_{b}(\dr + h_{00})+\partial_{t}h_{00}]  \nonumber \\
&-{1 \over 2}((3+4\epsilon)H_{b} \partial_{t} + \partial_{t}^{2})h_{ii} =0.
\end{align}

The fluid equations of motion in the long wavelength limit are,
\begin{align}
-{\d(\nabla_{\m} {T^{\m}}_0) \over \bar{\rho}} &= \partial_{t} \d_{\rho} +{1\over 2}(w+1) \partial_{t}h_{ii}=0, \\ 
{\d(\nabla_{\m} {T^{\m}}_i) \over (1+w)\bar{\rho} a(t)} &= [(1+2\epsilon-w(3+4\epsilon))H_{b} + \partial_{t}]\duf^{i} = 0, \\ 
{\d(\nabla_{\m} {T^{\m}}_3) \over (1+w) \bar{\rho} b(t)} &=[(1-w(3+4\epsilon))H_{b} +\partial_{t}]\duf^{3}=0.
\end{align}

And the equations of motion for the vector field are,
\begin{multline}
{\d\left(P_{0 \sigma}  \nabla_{\rho} F^{\rho\sigma}\right) \over m } = H_b^2 \left[ {1 - 3 w \over 2} + 2 \epsilon ( 1 - w) \right]\vf^{0}= 0,
\end{multline}
\vspace{-.5cm}
\begin{multline}
{\d\left(P_{i \sigma}  \nabla_{\rho} F^{\rho\sigma}\right) \over m a(t)} = 
\big[ H_b^2 \epsilon w (3 + 4 \epsilon) - \partial_t^2 \big](\vf^i+h_{i3}) \\
+ \epsilon H_b^2(\vf^i-h_{i3}) - (3 + 4 \epsilon)H_b \partial_t \vf^i \\ - (2 + 4 \epsilon)H_b \partial_t h_{i3} = 0,
 \end{multline}
 where $ P_{\m \sigma} \equiv g_{\m \sigma} -{u_\m u_{\sigma} \over m^{2} } $.

The coupled fields fall into two uncoupled sets  $\{\d_{\rho}, h_{00}, h_{ii} \}$ and $\{\vf^{i}, h_{i3}\}$, while the remaining degrees of freedom are uncoupled.  The solutions take the form, 
\begin{align}
\dr (t)&= -c_{\rho} - c_{00}{1 + w \over 2 w} (a^{2} b)^{(w-1)/2}, \\
h_{00}(t) &= c_{\rho} + c_{00}  (a^{2} b)^{(w-1)/2}, \\
h_{11}(t) &= h_{22}(t) + c_{22} = -{1 \over 1 + w} \dr(t) + c_{11}, \\
h_{12}(t)&= c_{12} + \tilde{c}_{12}(a^{2}b)^{{w-1 \over 2}} , \\
h_{i3}(t)&= {a \over b} \left(c_{i3}^{(1)} + c_{i3}^{(2)}(a^{2}b)^{{w-1 \over 2}}\right) \nonumber \\
&+ {b \over a} \left(c_{i3}^{(3)} + c_{i3}^{(4)}(a^{2}b)^{{w-1 \over 2}}\right), \label{hi3Solution} \\
\vf^{i}(t)&= {a \over b} \left(-c_{i3}^{(1)} +\tilde{c}_{i3}^{(2)}(a^{2}b)^{{w-1 \over 2}}\right) + \tilde{c}_{i3}^{(4)}{b \over a} (a^{2}b)^{{w-1 \over 2}}, \label{uSolution}
\end{align}
where,
\begin{align}
8\epsilon \tilde{c}_{i3}^{(2)}&= c_{i3}^{(2)}(6-(3+4\epsilon)(w+1)), \\
4\epsilon \tilde{c}_{i3}^{(4)}&=- c_{i3}^{(4)} {(w-1)(3+4\epsilon)(1+4\epsilon) \over  ((3+4\epsilon)(w-1)+2) }.
\end{align}
The constants $\{c_{\rho}, c_{00},...\}$ are fixed by initial conditions. The decaying parts of all comoving perturbations except $h_{i3}$ and $\vf^{i}$ go as  $(a^{2}b)^{{w-1 \over 2}}$.  Special attention must be paid to $h_{i3}$ and $\vf^{i}$ because each has a term proportional to $a/b = t^{4\epsilon/(3+4\epsilon)(1+w)}$  which grows with time.  

Consider a gauge transformation of the form ($x^{\m} \rightarrow x^{\m} + \xi^{\mu}$) where,
\begin{equation}
\xi_{0} = \xi_{3} =0 \qquad \text{and} \qquad \xi_{i}(t,z) = c_{i} z\, a(t)^{2}.
\end{equation}  
By Eqs.~\eqref{uGauge}-\eqref{metricGauge}, this gauge transformation leaves all perturbations except $h_{i3}$ and $\vf^{i}$ unchanged.  In particular the modified synchronous gauge conditions remain undisturbed. The gauge transformation simultaneously changes $\vf^{i}$ and $h_{i3}$ by,
\begin{align}
\Delta h_{i3} = -c_{i} {a\over b} ~~\text{and}~~ \Delta \vf^{i} =  c_{i} {a \over b}.  
\end{align}
Thus we see that the growing part of both $\vf^{i}$ and $h_{i3}$ can be gauged away because of the relative sign of the terms multiplying $a/b$ in~\eqref{hi3Solution} and~\eqref{uSolution}. 

Therefore, for $w < 1$ (true for ordinary matter) all modes with wavelengths larger than the Hubble radius decay to a constant.  In particular, $h_{i3}$ and $\vf^{i}$ decay to zero on a characteristic time scale proportional to $1/\epsilon$.

We recover the generic behavior of flat FRW perturbations in the long wavelength limit. As stated at the beginning of the section, FRW perturbations on scales much larger than the horizon have constant and decaying parts. 

For good measure, let us consider the exact correspondence in the $w=0$ case. For nonrelativistic, pressureless ($w=0$) matter, the metric is given by \eqref{bkgnd metric} with, 
\begin{equation}
a = b \propto \eta^2 \propto t^{2/3},
\end{equation}
where $\eta = \int dt / a$ is the conformal time. One can parameterize scalar perturbations to the FRW metric by \cite{Mukhanov},
\begin{multline}\label{isotropic perturbed metric}
ds^2 = a(\eta)^2 \Big[ -(1 + 2 \phi) d\eta^2 - 2 B_{, m} dx^m d\eta  + \\
	\left((1 - 2 \psi) \delta_{m n} - 2 E_{, m n} \right) dx^m dx^n \Big],
\end{multline}
where $m,n \in \{ 1, 2, 3\}$. Then the function,
\begin{equation}
\Phi = \phi - {1 \over a}(a(B - E'))',
\end{equation}
where $'$ denotes ${\partial \over \partial \eta}$, is gauge invariant. The relationship between the comoving  perturbations, $h_{\m \n}$,  defined in \eqref{perturbed metric} and the perturbations defined in \eqref{isotropic perturbed metric} in Fourier space is,
\begin{align}
\psi  &= {1 \over 2} \left( {k_m k_n \over \vec{k}^2} - \delta_{m n} \right)h_{m n}, \\
E &= {1 \over 4 \vec{k}^2} \left( 3 { k_m k_n \over \vec{k}^2 } - \delta_{m n} \right) h_{m n},   \\
B & = i {k_n \over \vec{k}^2} h_{0 n}, \qquad \phi = { h_{00} \over 2}.
\end{align}
In our modified synchronous gauge, $B = 0$. So, in the isotropic limit ($b \rightarrow a$) when $w=0$, we find the Fourier transformed gauge invariant function to be,
\begin{align}
\tilde{\Phi} 
	&= {h_{00} \over 2} - {1 \over 4 \vec{k}^2} \left( 3 { k_m k_n \over \vec{k}^2 } - \delta_{m n} \right)({a' \over a} h_{m n}' + h_{m n}''), \nonumber \\
	&= {c_{\rho} \over 2} + { c_{00} \over 2} {\eta_0^3 \over \eta^3} - {1 \over 4 \vec{k}^2} {\eta_0^3 \over \eta^5} \left( 3 { k_m k_n \over \vec{k}^2 } - \delta_{m n} \right) \nonumber \\
	&~~~~\times(2 \eta_0 h_{m n}'(\eta_0) + \eta_0^2 h_{m n}''(\eta_0)), \nonumber \\
	&= {c_\rho \over 2} +  {\eta_0^3 \over \eta^3}\Big[  { c_{00} \over 2} - { |\vec{k} \eta|^{-2} \over 4}  \left( 3 { k_m k_n \over \vec{k}^2 } - \delta_{m n} \right) \nonumber \\
	&~~~~\times(2 \eta_0 h_{m n}'(\eta_0) + \eta_0^2 h_{m n}''(\eta_0))\Big],
\end{align}
since $a \propto \eta^2$. Here we've defined $a(\eta_0) = 1$. In the long wavelength limit, $|\vec{k} \eta| << 1$, and so the second term in the brackets dominates over the first for all times. Thus we have, in the long wavelength limit,
\begin{equation}
\Phi \sim C_1(\vec{x}) + C_2 (\vec{x}) \eta^{-5}.
\end{equation}
This is precisely the functional form found for $\Phi$ in a $w=0$ flat FRW universe \cite{Mukhanov}. One can work out similar correspondences for other gauge invariant quantities in the isotropic, long wavelength limit.

%%%%%%%%%%%%%%%%%%%%%%%%%%%%%%%%%%%%%%%%%

\section{Observational limits}

The isotropy of the microwave background was used to place present-day limits on the density of matter with general anisotropic stresses in \cite{Barrow:1997sy}.\footnote{An analysis of the of the cosmological evolution of matter sources with small anisotropic pressures in various cosmological epochs was carried out in \cite{Barrow:1997sy} An anisotropic equation of state for dark energy was considered in \cite{Koivisto:2007bp}.}.  If we assume that the VEV of the vector field, $m$, was constant from the time of recombination, then, following \cite{Barrow:1997sy}, we may calculate the temperature anisotropy due to the direction-dependent evolution of the photon temperature since recombination,
\begin{align}
T_{x_{1}} = T_{x_{2}} &= T_0 { a_0 \over a(t) } = T_0 \exp \left[ - \int_{t_0}^t H_a dt \right] \\
T_{x_{3}} &= T_0 {b_0 \over b(t) } = T_0 \exp \left[ - \int_{t_0}^t H_b dt \right],
\end{align}
by,
\begin{align}
{\d T \over T} & \equiv {T_{x_{1}} - T_{x_{3}} \over T_0 } \\ \nonumber
	&=  \exp \left[ -(1 + 2 \epsilon ) \int_{t_0}^t H_b dt \right] - \exp \left[ - \int_{t_0}^t H_b dt \right]\\ \nonumber 
	&\approxeq 2 \epsilon \left( - \int_{t_0}^t H_b dt \right) \exp \left[ -\int_{t_0}^t H_b dt \right] \\
	&\approxeq 2 \epsilon {a_0 \over a(t)}  \log \left( {a_0 \over a(t)} \right).
\end{align}
Thus,
\begin{equation}
{\d T \over T} \big|_\text{today} \approxeq 2 \epsilon (1 + z_{rec})  \log (1 + z_{rec}).
\end{equation}

Observations show that \cite{PDBook},
\begin{equation}
{\d T \over T} \leq 10^{-5}.
\end{equation}
Taking $1 + z_{rec} \approxeq 1100$ then we have the limit on $\epsilon = 4 \pi G m^2$,
\begin{equation}
2 \epsilon \lesssim 10^{-9}.
\end{equation}
Plugging in $(8 \pi G)^{-1/2} = 2.43 \times 10^{18} GeV$ gives the limit,
\begin{equation} \label{mlimit}
m \lesssim  10^{14} GeV .
\end{equation}

Presence of the vector field during inflation also affects the CMB.  The analysis in \cite{Pullen:2007tu}, indicates that the parameter defined in  \cite{Ackerman:2007}, $g_*$, characterizing deviation of the power spectrum from isotropy during the inflationary era,
\begin{equation}
P(\mathbf{k}) = \bar{P}(|\mathbf{k}|) \left( 1 + g_* (\hat{\mathbf{k}} \cdot \mathbf{n} )^2 \right),
\end{equation}
can be measured by the Planck experiment down to a value of (roughly) $10^{-2}$. The parameter $g_*$ is roughly equal to $\epsilon = 4 \pi G m^2$ \cite{Ackerman:2007}.  Thus, the vector field could have had a much larger VEV during inflation than during recombination (and thereafter) and still escape detection by Planck. If, however, the vector field has a constant VEV throughout cosmological history, the limit~\eqref{mlimit} indicates that the effect of the vector field during inflation will be undetectable.

Bounds from primordial nucleosynthesis were also considered in \cite{Barrow:1997sy}. Primoridial nucleosynthesis limits the change to the mean expansion rate imposed at the epoch of neutron-proton freeze-out. If the rate of cosmic expansion during Big Bang nucleosynthesis is slower, then weak interactions freeze out later, at a lower temperature. A lower temperature at freeze-out results in a decrease in the primoridal ${}^{4}$He to H mass ratio, which is constrained by observation. 
It was noted in \cite{Barrow:1997sy} that if the anisotropy does not decay (or decays very slowly) between the period of freeze-out and matter-radiation equality, then bounds from the temperature anisotropy are actually stronger than bounds from primoridal nucleosynthesis. For example, a bound from primordial nucleosynthesis (due to the difference between the effective Newton's constant measured in the solar system versus the effective cosmic expansion Newton's constant) on the VEV of a similar timelike vector field was found; for a normalization similar to ours, \cite{Carroll:2004ai} found $m < 10^{18} GeV$.

It would also be interesting to consider how a fixed-norm spacelike vector field modifies gravity in the weak field limit. Furthermore, it would be interesting to see whether the direction-dependent Jeans instability that we found significantly affects structure formation.

%%%%%%%%%%%%%%%%%%%%%%%%%%%%%%%%%%%%%%%%%

\section{Concluding remarks}

In this paper, we considered an anisotropic Bianchi type I space-time with energy density that is dominated by that of a perfect fluid with equation of state $p = w \rho$ and whose anisotropy is due to a fixed-norm spacelike vector field.  We derived the background space-time (Eqs.~\eqref{bkgnd metric},~\eqref{Ha} and~\eqref{Lambda}),
\begin{align}
{\rm d}s^2 &=-{\rm d}t^2+a(t)^2{\rm d}{\bf x}_{\perp}^2+b(t)^2{\rm d}x_3^2, \\
H_a(t) &= {{\dot a}\over a} = H_b(t)(1+2 \epsilon ),   \\
H_b(t) &= {{\dot b}\over b} = \sqrt{{8\pi G \bar{\rho}(t) \over (1 +  \epsilon)(3 + 4 \epsilon)}},
\end{align}
where $\epsilon \equiv 4\pi G m^{2}$.

We went on to consider first order perturbations about the background space-time, assuming negligible shear-stress, and found that the space-time is as perturbatively stable as the flat FRW universe.  We derived an anisotropic generalization of the Jeans instability for an anisotropic $w=0$ FRW-like universe. We showed that energy density perturbations parallel to the direction of the vector field VEV are governed by Eq.~\eqref{k3Jeans},
\begin{equation}
\ddot{\d}_{\rho} + 2 H_{b} \dot{\d}_{\rho} = (4\pi G \bar{\rho}) \d_{\rho},
\end{equation}
while the perturbations perpendicular to the VEV are governed by Eq.~\eqref{kpJeans},
\begin{equation} 
\ddot{\d}_{\rho} + 2H_{a}\dot{\d}_{\rho}= \left({4\pi G \bar{\rho} /(1+\epsilon)}\right) \d_{\rho}.
\end{equation}
These equations imply that the (parallel/perpendicular) perturbations grow at a (faster/slower) rate than the usual Jeans instability predicts for a flat FRW universe.

We computed the rough limit, $m \lesssim  10^{14} GeV$, assuming the vector field was present during recombination and thereafter.  The effects of the vector field, if present only during inflation, are much harder to detect.   Stronger bounds might arise from analyzing solar system tests of gravity, structure formation, or SN1a redshift data.  

%%%%%%%%%%%%%%%%%%%%%%%%%%%%%%%%%%%%%%%%%

\section{Acknowledgments}

We thank Sean Carroll, Matt Johnson and Mark Wise for invaluable conversations during the preparation of this manuscript.  

%%%%%%%%%%%%%%%%%%%%%%%%%%%%%%%%%%%%%%%%%
\appendix
\section{Perturbation equations}
We Fourier transform the metric perturbations in space,
\begin{align}
h_{\m \n}(t, \vec{k}) = \int d^3 x\, h_{\m \n}(t, \vec{x})  e^{i \vec{k} \cdot \vec{x}},
\end{align}
as well as the energy density contrast, vector field and fluid velocity perturbations. Because the background space-time is spatially homogenous, perturbations with different comoving wavenumbers will evolve independently.  The physical wavenumbers $\bar{k}_{i,3}$ are defined in terms of the comoving wavenumbers by,
\begin{equation}
\bar{k}_{i} = {k_{i} \over a(t)} \qquad \text{and} \qquad \bar{k}_{3} = {k_{3} \over b(t)}.
\end{equation}

The following are the Fourier transformed first-order field equations, $\delta E_{\m\n}=\delta G_{\m\n}- 8\pi G \delta T_{\m\n}=0$, for the metric, vector, and fluid perturbations defined in~Eqs.\eqref{perturbed metric}-\eqref{perturbed v}.  We use the background solution of Eq.~\eqref{Ha} and Eq.~\eqref{Lambda} to eliminate $H_{a}$ and $\bar{\rho}$.  In these equations, repeated $\{i,j,k,l\}$ indices are summed over $\{1,2\}$ and $\vec{\bar{k}}^{2} \equiv \kpb^{2} + \bar{k}_{3}^{2}$.
\begin{widetext}
\begin{align} \label{e00}
\delta E_{00} &= -(3+4\epsilon)(1+\epsilon) H_{b}^{2}(\delta_{\rho}+h_{00}) - 2 i \epsilon H_{b} \bar{k}_{3} \vf^{0} -2 (1+\epsilon)H_{b} \left( i \bar{k}_{j} h_{0j} + i \bar{k}_{3} h_{03} \right) + {1 \over 2} (\d_{ij} \vec{\bar{k}}^{2} -\bar{k}_{i}\bar{k}_{j})h_{ij} \nonumber \\
&-\bar{k}_{j} \bar{k}_{3}h_{j3} +(1+\epsilon) H_{b}\partial_{t} h_{jj} + {1 \over 2}( \kpb^{2} + 2(1+\epsilon)H_{b} \partial_{t}) h_{33} =0, 
\end{align}
\begin{align}
{\delta E_{0i} \over a(t)} &= (1+w)(3+4\epsilon)(1+\epsilon)H_{b}^{2}(\duf^{i} +h_{0i}) + 2 i \epsilon H_{b} \bar{k}_{3} \vf^{i}+(1+\epsilon)H_{b} i \bar{k}_{i} h_{00}+ {1 \over 2}(\d_{ij} \vec{\bar{k}}^{2} -\bar{k}_{i} \bar{k}_{j})h_{0j} \nonumber \\ 
&- {1 \over 2} \bar{k}_{i} \bar{k}_{3} h_{03} + {1 \over 2}  i \bar{k}_{3}  \left( 2 \epsilon H_{b} +  \partial_{t} \right) h_{i3} +{1 \over 2}( i \bar{k}_{j} \partial_{t} h_{ij} - i \bar{k}_{i} \partial_{t}h_{jj})  -{1\over 2} i \bar{k}_{i}\partial_{t}h_{33}=0, 
\end{align}
\begin{align}
{\delta E_{03} \over b(t)} &= (1+w)(3+4\epsilon)(1+\epsilon)H_{b}^{2}(\duf^{3} +h_{03}) - 2\epsilon \vec{\bar{k}}^{2} \vf^{0} + 2\epsilon (1 + 2\epsilon) i \bar{k}_{i} \vf^{i}+ 2 \epsilon i \bar{k}_{i} \partial_{t}\vf^{i}+(1+\epsilon)H_{b} i \bar{k}_{3} h_{00} \nonumber \\
&- {1 \over 2} \bar{k}_{i} \bar{k}_{3} h_{0i} +{1 \over 2}((1+4\epsilon) \vec{\bar{k}}^{2} - \bar{k}_{3}^{2})h_{03} +{1 \over 2}(1+4\epsilon)i \bar{k}_{j}(2\epsilon H_{b} +\partial_{t}) h_{j3} - {1 \over 2} i \bar{k}_{3} \partial_{t} h_{ii} + \epsilon i \k3 \partial_{t} h_{33}=0,  
\end{align}
\begin{align} \label{eij}
{\delta E_{ij} \over a(t)^{2}} &= - w (1 + \epsilon)(3 + 4 \epsilon) H_b^2 \d_{i j} (\d_\r + h_{0 0}) -  2 i \epsilon \delta_{ij} H_{b} \bar{k}_{3} \vf^{0} + {1 \over 2} \left( \ki \kj  - \d_{ij}\left(\vec{\bar{k}}^{2} - 2 (1 + \epsilon) H_b \partial_t \right) \right) h_{00}  \nonumber \\
	& + i \k3 \d_{i j} \left( 2(1 + 2 \epsilon) H_b + \partial_t \right) h_{03} - \e_{l ( i} \e_{j ) k} \left( i \bar{k}_l \left( 2(1 + \epsilon) H_b + \partial_t \right) h_{0 k} + \k3 \bar{k}_l h_{k 3} \right) \nonumber \\
		& + {1 \over 2} \e_{l ( i} \e_{j ) k} \left( \k3^2 + (3 + 4 \epsilon) H_b \partial_t + \partial_t^2 \right)h_{l k} +{1 \over 2} (\bar{k}_{i}\bar{k}_{j} - \delta_{ij} \kpb^{2}) h_{33} - \delta_{ij}{1 \over 2}( (3 + 4\epsilon) H_{b}\partial_{t} +\partial_{t}^{2}) h_{33}= 0,
\end{align}
\begin{align}
{2\delta E_{i3} \over a(t)b(t)} &=- i \bar{k}_{3}(2H_{b} (1+\epsilon) +\partial_{t}) h_{0i} - (1+4\epsilon) i \bar{k}_{i} (2 H_{b}+\partial_{t})h_{03} -\bar{k}_{3}( \bar{k}_{j} h_{ij} - \bar{k}_{i} h_{jj})  + \bar{k}_{i}\bar{k}_{3} h_{00}  - \bar{k}_{3}^{2} h_{i3} -2\epsilon \bar{k}_{i} \bar{k}_{3} h_{33}  \nonumber \\
& + (1+4 \epsilon)(  \vec{\bar{k}}^{2} \d_{ij}-\bar{k}_{i}\bar{k}_{j})h_{j3} + (1+4\epsilon)\left[ ( 3 - w(3 + 4 \epsilon) ) \epsilon H_{b}^{2} + (3 + 4 \epsilon) H_{b}\partial_{t} + \partial_{t}^{2} \right] h_{i3} + 4\epsilon i \bar{k}_{i}(2H_{b}+ \partial_{t}) \vf^{0} \nonumber \\
& + 2\epsilon \left( (1+2\epsilon)(3-w(3+4\epsilon))H_{b}^{2} +  8(1+\epsilon)H_{b} \partial_{t} + 2 \partial_{t}^{2} \right) \vf^{i} + 4\epsilon (\delta_{ij} \vec{\bar{k}}^{2} - \bar{k}_{i} \bar{k}_{j}) \vf^{j} =0,
\end{align}
\begin{align}
{\delta E_{33} \over b(t)^{2}} &=-w(3+4\epsilon)(1+\epsilon)H_{b}^{2}(\d_{\rho}+h_{00})- {1\over 2} \kpb^{2}h_{00} + {1 \over 2} (\bar{k}_{i}\bar{k}_{j}-\kpb^{2}\d_{ij})h_{ij} -2\epsilon \bar{k}_{i}\bar{k}_{3} h_{i3}+ i \bar{k}_{i}(2(1+\epsilon)H_{b} + \partial_{t}) h_{0i} \nonumber \\
&-2i \bar{k}_{3}\epsilon(2(1+2\epsilon)H_{b} + \partial_{t}) h_{03}  +(1+\epsilon)H_{b} \partial_{t}h_{00}-{1 \over 2}((3+4\epsilon)H_{b} \partial_{t} + \partial_{t}^{2})h_{ii} +4\epsilon(1+2\epsilon)H_{b}i \bar{k}_{3} \vf^{0} - 2\epsilon \bar{k}_{i} \bar{k}_{3} \vf^{i} \nonumber \\
& + \epsilon \kpb^{2} h_{33} +2i\epsilon k_{3} \partial_{t} \vf^{0} + \epsilon((3+4\epsilon)H_{b} \partial_{t} + \partial_{t}^{2})h_{33}=0.
\end{align}

We also have the following equations resulting from the fluid equations of motion,
\begin{align}
{\d(\nabla_{\m} {T^{(fl)}{}^{\m}}_0) \over \bar{\rho}} &= -(1+w) i \bar{k}_{i} \duf^{i} -(1+w)i \bar{k}_{3} \duf^{3} -\partial_{t} \d_{\rho} -{1\over 2}(w+1) \partial_{t}(h_{ii} + h_{33})=0, \\ \label{fluid ti}
{\d(\nabla_{\m} {T^{(fl)}{}^{\m}}_i) \over \bar{\rho} a(t)} &= w i\bar{k}_{i}\d_{\rho} -(1+w)((3w-1 +2(2w-1)\epsilon)H_{b} - \partial_{t})(\duf^{i}+h_{0i}) +{1\over 2}(1+w)i\bar{k}_{i} h_{00}, \\ \label{fluid t3}
{\d(\nabla_{\m} {T^{(fl)}{}^{\m}}_3) \over \bar{\rho} b(t)} &= w i\bar{k}_{3}\d_{\rho} -(1+w)((3w-1+4w\epsilon)H_{b} -\partial_{t})(\duf^{3}+h_{03}) +{1\over 2}(1+w)i\bar{k}_{3} h_{00}.
\end{align}

The equations of motion for the vector field are ($ P_{\m \sigma} \equiv g_{\m \sigma} -{u_\m u_{\sigma} \over m^{2} } $):
\begin{align}
{ \delta \left(P_{0 \sigma} \nabla_{\rho} F^{\rho\sigma} \right) \over m } 
	&=  - H_b^2 \left( {1 \over 2}(1 - 3 w) + 2 \epsilon ( 1 - w) \right)(\vf^{0}) + (\kpb^2 + \k3^2)( \vf^0 - h_{03} ) \\
	& - i \ki ((1 + 2 \epsilon) H_b + \partial_t)(\vf^i + h_{i 3}) + {1 \over 2} H_b i \k3( h_{00}- h_{ii}) - {1 \over 2} i \k3 \partial_t h_{33} + H_b i \ki h_{i 3}
	= 0, \\
{ \delta \left( P_{i \sigma}\nabla_{\rho} F^{\rho\sigma} \right) \over m a(t) } 
	&= - i \ki (H_b + \partial_t)(\vf^0- h_{03}) + \ki \kj (\vf^j +h_{j 3}) -  i \k3 H_b h_{0i} -(2 \epsilon H_b^2 - H_b \partial_t)h_{i 3} \nonumber \\
	&+ {1 \over 2} \ki \k3 h_{33} + ( H_b^2 \epsilon (1 + w (3 + 4 \epsilon)) - \kpb^2 - \k3^2 -(3 + 4 \epsilon)H_b \partial_t - \partial_t^2)(\vf^i  + h_{i 3})
		= 0.
\end{align}
\end{widetext}

%%%%%%%%%%%%%%%%%%%%%%%%%%%%%%%%%%%%%%%%%

\bibliography{references}

\begin{thebibliography}{44}
\expandafter\ifx\csname natexlab\endcsname\relax\def\natexlab#1{#1}\fi
\expandafter\ifx\csname bibnamefont\endcsname\relax
  \def\bibnamefont#1{#1}\fi
\expandafter\ifx\csname bibfnamefont\endcsname\relax
  \def\bibfnamefont#1{#1}\fi
\expandafter\ifx\csname citenamefont\endcsname\relax
  \def\citenamefont#1{#1}\fi
\expandafter\ifx\csname url\endcsname\relax
  \def\url#1{\texttt{#1}}\fi
\expandafter\ifx\csname urlprefix\endcsname\relax\def\urlprefix{URL }\fi
\providecommand{\bibinfo}[2]{#2}
\providecommand{\eprint}[2][]{\url{#2}}

\bibitem[{\citenamefont{Kostelecky and Lane}(1999)}]{Kostelecky:1999mr}
\bibinfo{author}{\bibfnamefont{V.~A.} \bibnamefont{Kostelecky}}
  \bibnamefont{and} \bibinfo{author}{\bibfnamefont{C.~D.} \bibnamefont{Lane}},
  \bibinfo{journal}{Phys. Rev.} \textbf{\bibinfo{volume}{D60}},
  \bibinfo{pages}{116010} (\bibinfo{year}{1999}), \eprint{hep-ph/9908504}.

\bibitem[{\citenamefont{Bluhm et~al.}(2000)\citenamefont{Bluhm, Kostelecky, and
  Lane}}]{Bluhm:1999dx}
\bibinfo{author}{\bibfnamefont{R.}~\bibnamefont{Bluhm}},
  \bibinfo{author}{\bibfnamefont{V.~A.} \bibnamefont{Kostelecky}},
  \bibnamefont{and} \bibinfo{author}{\bibfnamefont{C.~D.} \bibnamefont{Lane}},
  \bibinfo{journal}{Phys. Rev. Lett.} \textbf{\bibinfo{volume}{84}},
  \bibinfo{pages}{1098} (\bibinfo{year}{2000}), \eprint{hep-ph/9912451}.

\bibitem[{\citenamefont{Bluhm and Kostelecky}(2000)}]{Bluhm:1999ev}
\bibinfo{author}{\bibfnamefont{R.}~\bibnamefont{Bluhm}} \bibnamefont{and}
  \bibinfo{author}{\bibfnamefont{V.~A.} \bibnamefont{Kostelecky}},
  \bibinfo{journal}{Phys. Rev. Lett.} \textbf{\bibinfo{volume}{84}},
  \bibinfo{pages}{1381} (\bibinfo{year}{2000}), \eprint{hep-ph/9912542}.

\bibitem[{\citenamefont{Kostelecky and Mewes}(2001)}]{Kostelecky:2001mb}
\bibinfo{author}{\bibfnamefont{V.~A.} \bibnamefont{Kostelecky}}
  \bibnamefont{and} \bibinfo{author}{\bibfnamefont{M.}~\bibnamefont{Mewes}},
  \bibinfo{journal}{Phys. Rev. Lett.} \textbf{\bibinfo{volume}{87}},
  \bibinfo{pages}{251304} (\bibinfo{year}{2001}), \eprint{hep-ph/0111026}.

\bibitem[{\citenamefont{Kostelecky and Mewes}(2002)}]{Kostelecky:2002hh}
\bibinfo{author}{\bibfnamefont{V.~A.} \bibnamefont{Kostelecky}}
  \bibnamefont{and} \bibinfo{author}{\bibfnamefont{M.}~\bibnamefont{Mewes}},
  \bibinfo{journal}{Phys. Rev.} \textbf{\bibinfo{volume}{D66}},
  \bibinfo{pages}{056005} (\bibinfo{year}{2002}), \eprint{hep-ph/0205211}.

\bibitem[{\citenamefont{Cane et~al.}(2004)}]{Cane:2003wp}
\bibinfo{author}{\bibfnamefont{F.}~\bibnamefont{Cane}} \bibnamefont{et~al.},
  \bibinfo{journal}{Phys. Rev. Lett.} \textbf{\bibinfo{volume}{93}},
  \bibinfo{pages}{230801} (\bibinfo{year}{2004}), \eprint{physics/0309070}.

\bibitem[{\citenamefont{Bailey and Kostelecky}(2004)}]{Bailey:2004na}
\bibinfo{author}{\bibfnamefont{Q.~G.} \bibnamefont{Bailey}} \bibnamefont{and}
  \bibinfo{author}{\bibfnamefont{V.~A.} \bibnamefont{Kostelecky}},
  \bibinfo{journal}{Phys. Rev.} \textbf{\bibinfo{volume}{D70}},
  \bibinfo{pages}{076006} (\bibinfo{year}{2004}), \eprint{hep-ph/0407252}.

\bibitem[{\citenamefont{Moore and Nelson}(2001)}]{Moore:2001bv}
\bibinfo{author}{\bibfnamefont{G.~D.} \bibnamefont{Moore}} \bibnamefont{and}
  \bibinfo{author}{\bibfnamefont{A.~E.} \bibnamefont{Nelson}},
  \bibinfo{journal}{JHEP} \textbf{\bibinfo{volume}{09}}, \bibinfo{pages}{023}
  (\bibinfo{year}{2001}), \eprint{hep-ph/0106220}.

\bibitem[{\citenamefont{Burgess et~al.}(2002)\citenamefont{Burgess, Cline,
  Filotas, Matias, and Moore}}]{Burgess:2002tb}
\bibinfo{author}{\bibfnamefont{C.~P.} \bibnamefont{Burgess}},
  \bibinfo{author}{\bibfnamefont{J.}~\bibnamefont{Cline}},
  \bibinfo{author}{\bibfnamefont{E.}~\bibnamefont{Filotas}},
  \bibinfo{author}{\bibfnamefont{J.}~\bibnamefont{Matias}}, \bibnamefont{and}
  \bibinfo{author}{\bibfnamefont{G.~D.} \bibnamefont{Moore}},
  \bibinfo{journal}{JHEP} \textbf{\bibinfo{volume}{03}}, \bibinfo{pages}{043}
  (\bibinfo{year}{2002}), \eprint{hep-ph/0201082}.

\bibitem[{\citenamefont{Kostelecky}(2004)}]{Kostelecky:2003fs}
\bibinfo{author}{\bibfnamefont{V.~A.} \bibnamefont{Kostelecky}},
  \bibinfo{journal}{Phys. Rev.} \textbf{\bibinfo{volume}{D69}},
  \bibinfo{pages}{105009} (\bibinfo{year}{2004}), \eprint{hep-th/0312310}.

\bibitem[{\citenamefont{Bailey and Kostelecky}(2006)}]{Bailey:2006fd}
\bibinfo{author}{\bibfnamefont{Q.~G.} \bibnamefont{Bailey}} \bibnamefont{and}
  \bibinfo{author}{\bibfnamefont{V.~A.} \bibnamefont{Kostelecky}},
  \bibinfo{journal}{Phys. Rev.} \textbf{\bibinfo{volume}{D74}},
  \bibinfo{pages}{045001} (\bibinfo{year}{2006}), \eprint{gr-qc/0603030}.

\bibitem[{\citenamefont{Kostelecky and Samuel}(1989)}]{Kostelecky:1989jw}
\bibinfo{author}{\bibfnamefont{V.~A.} \bibnamefont{Kostelecky}}
  \bibnamefont{and} \bibinfo{author}{\bibfnamefont{S.}~\bibnamefont{Samuel}},
  \bibinfo{journal}{Phys. Rev.} \textbf{\bibinfo{volume}{D40}},
  \bibinfo{pages}{1886} (\bibinfo{year}{1989}).

\bibitem[{\citenamefont{Jacobson and Mattingly}(2004)}]{Jacobson:2004ts}
\bibinfo{author}{\bibfnamefont{T.}~\bibnamefont{Jacobson}} \bibnamefont{and}
  \bibinfo{author}{\bibfnamefont{D.}~\bibnamefont{Mattingly}},
  \bibinfo{journal}{Phys. Rev.} \textbf{\bibinfo{volume}{D70}},
  \bibinfo{pages}{024003} (\bibinfo{year}{2004}), \eprint{gr-qc/0402005}.

\bibitem[{\citenamefont{Carroll and Lim}(2004)}]{Carroll:2004ai}
\bibinfo{author}{\bibfnamefont{S.~M.} \bibnamefont{Carroll}} \bibnamefont{and}
  \bibinfo{author}{\bibfnamefont{E.~A.} \bibnamefont{Lim}},
  \bibinfo{journal}{Phys. Rev.} \textbf{\bibinfo{volume}{D70}},
  \bibinfo{pages}{123525} (\bibinfo{year}{2004}), \eprint{hep-th/0407149}.

\bibitem[{\citenamefont{Lim}(2005)}]{Lim:2004js}
\bibinfo{author}{\bibfnamefont{E.~A.} \bibnamefont{Lim}},
  \bibinfo{journal}{Phys. Rev.} \textbf{\bibinfo{volume}{D71}},
  \bibinfo{pages}{063504} (\bibinfo{year}{2005}), \eprint{astro-ph/0407437}.

\bibitem[{\citenamefont{Eling and Jacobson}(2004)}]{Eling:2003rd}
\bibinfo{author}{\bibfnamefont{C.}~\bibnamefont{Eling}} \bibnamefont{and}
  \bibinfo{author}{\bibfnamefont{T.}~\bibnamefont{Jacobson}},
  \bibinfo{journal}{Phys. Rev.} \textbf{\bibinfo{volume}{D69}},
  \bibinfo{pages}{064005} (\bibinfo{year}{2004}), \eprint{gr-qc/0310044}.

\bibitem[{\citenamefont{Foster and Jacobson}(2006)}]{Foster:2005dk}
\bibinfo{author}{\bibfnamefont{B.~Z.} \bibnamefont{Foster}} \bibnamefont{and}
  \bibinfo{author}{\bibfnamefont{T.}~\bibnamefont{Jacobson}},
  \bibinfo{journal}{Phys. Rev.} \textbf{\bibinfo{volume}{D73}},
  \bibinfo{pages}{064015} (\bibinfo{year}{2006}), \eprint{gr-qc/0509083}.

\bibitem[{\citenamefont{Li et~al.}(2007)\citenamefont{Li, Fonseca~Mota, and
  Barrow}}]{Li:2007vz}
\bibinfo{author}{\bibfnamefont{B.}~\bibnamefont{Li}},
  \bibinfo{author}{\bibfnamefont{D.}~\bibnamefont{Fonseca~Mota}},
  \bibnamefont{and} \bibinfo{author}{\bibfnamefont{J.~D.} \bibnamefont{Barrow}}
  (\bibinfo{year}{2007}), \eprint{arXiv:0709.4581 [astro-ph]}.

\bibitem[{\citenamefont{Seifert}(2007)}]{Seifert:2007fr}
\bibinfo{author}{\bibfnamefont{M.~D.} \bibnamefont{Seifert}},
  \bibinfo{journal}{Phys. Rev.} \textbf{\bibinfo{volume}{D76}},
  \bibinfo{pages}{064002} (\bibinfo{year}{2007}), \eprint{gr-qc/0703060}.

\bibitem[{\citenamefont{Foster}(2006)}]{Foster:2006az}
\bibinfo{author}{\bibfnamefont{B.~Z.} \bibnamefont{Foster}},
  \bibinfo{journal}{Phys. Rev.} \textbf{\bibinfo{volume}{D73}},
  \bibinfo{pages}{104012} (\bibinfo{year}{2006}), \eprint{gr-qc/0602004}.

\bibitem[{\citenamefont{Eling and Jacobson}(2006)}]{Eling:2006df}
\bibinfo{author}{\bibfnamefont{C.}~\bibnamefont{Eling}} \bibnamefont{and}
  \bibinfo{author}{\bibfnamefont{T.}~\bibnamefont{Jacobson}},
  \bibinfo{journal}{Class. Quant. Grav.} \textbf{\bibinfo{volume}{23}},
  \bibinfo{pages}{5625} (\bibinfo{year}{2006}), \eprint{gr-qc/0603058}.

\bibitem[{\citenamefont{Jacobson and Mattingly}(2001)}]{Jacobson:2000xp}
\bibinfo{author}{\bibfnamefont{T.}~\bibnamefont{Jacobson}} \bibnamefont{and}
  \bibinfo{author}{\bibfnamefont{D.}~\bibnamefont{Mattingly}},
  \bibinfo{journal}{Phys. Rev.} \textbf{\bibinfo{volume}{D64}},
  \bibinfo{pages}{024028} (\bibinfo{year}{2001}), \eprint{gr-qc/0007031}.

\bibitem[{\citenamefont{Clayton}(2001)}]{Clayton:2001vy}
\bibinfo{author}{\bibfnamefont{M.~A.} \bibnamefont{Clayton}}
  (\bibinfo{year}{2001}), \eprint{gr-qc/0104103}.

\bibitem[{\citenamefont{Graesser et~al.}(2005)\citenamefont{Graesser, Jenkins,
  and Wise}}]{Graesser:2005bg}
\bibinfo{author}{\bibfnamefont{M.~L.} \bibnamefont{Graesser}},
  \bibinfo{author}{\bibfnamefont{A.}~\bibnamefont{Jenkins}}, \bibnamefont{and}
  \bibinfo{author}{\bibfnamefont{M.~B.} \bibnamefont{Wise}},
  \bibinfo{journal}{Phys. Lett.} \textbf{\bibinfo{volume}{B613}},
  \bibinfo{pages}{5} (\bibinfo{year}{2005}), \eprint{hep-th/0501223}.

\bibitem[{\citenamefont{Kanno and Soda}(2006)}]{Kanno:2006ty}
\bibinfo{author}{\bibfnamefont{S.}~\bibnamefont{Kanno}} \bibnamefont{and}
  \bibinfo{author}{\bibfnamefont{J.}~\bibnamefont{Soda}},
  \bibinfo{journal}{Phys. Rev.} \textbf{\bibinfo{volume}{D74}},
  \bibinfo{pages}{063505} (\bibinfo{year}{2006}), \eprint{hep-th/0604192}.

\bibitem[{\citenamefont{Bluhm et~al.}(2007)\citenamefont{Bluhm, Fung, and
  Kostelecky}}]{Bluhm:2007bd}
\bibinfo{author}{\bibfnamefont{R.}~\bibnamefont{Bluhm}},
  \bibinfo{author}{\bibfnamefont{S.-H.} \bibnamefont{Fung}}, \bibnamefont{and}
  \bibinfo{author}{\bibfnamefont{V.~A.} \bibnamefont{Kostelecky}}
  (\bibinfo{year}{2007}), \eprint{arXiv:0712.4119 [hep-th]}.

\bibitem[{\citenamefont{Jacobson}(2008)}]{Jacobson:2008aj}
\bibinfo{author}{\bibfnamefont{T.}~\bibnamefont{Jacobson}}
  (\bibinfo{year}{2008}), \eprint{arXiv:0801.1547 [gr-qc]}.

\bibitem[{\citenamefont{Barrow}(1997)}]{Barrow:1997sy}
\bibinfo{author}{\bibfnamefont{J.~D.} \bibnamefont{Barrow}},
  \bibinfo{journal}{Phys. Rev.} \textbf{\bibinfo{volume}{D55}},
  \bibinfo{pages}{7451} (\bibinfo{year}{1997}), \eprint{gr-qc/9701038}.

\bibitem[{\citenamefont{Ackerman et~al.}(2007)\citenamefont{Ackerman, Carroll,
  and Wise}}]{Ackerman:2007}
\bibinfo{author}{\bibfnamefont{L.}~\bibnamefont{Ackerman}},
  \bibinfo{author}{\bibfnamefont{S.~M.} \bibnamefont{Carroll}},
  \bibnamefont{and} \bibinfo{author}{\bibfnamefont{M.~B.} \bibnamefont{Wise}},
  \bibinfo{journal}{Phys. Rev.} \textbf{\bibinfo{volume}{D75}},
  \bibinfo{pages}{083502} (\bibinfo{year}{2007}), \eprint{astro-ph/0701357}.

\bibitem[{\citenamefont{Gumrukcuoglu et~al.}(2007)\citenamefont{Gumrukcuoglu,
  Contaldi, and Peloso}}]{Gumrukcuoglu:2007bx}
\bibinfo{author}{\bibfnamefont{A.~E.} \bibnamefont{Gumrukcuoglu}},
  \bibinfo{author}{\bibfnamefont{C.~R.} \bibnamefont{Contaldi}},
  \bibnamefont{and} \bibinfo{author}{\bibfnamefont{M.}~\bibnamefont{Peloso}},
  \bibinfo{journal}{JCAP} \textbf{\bibinfo{volume}{0711}}, \bibinfo{pages}{005}
  (\bibinfo{year}{2007}), \eprint{arXiv:0707.4179 [astro-ph]}.

\bibitem[{\citenamefont{Gumrukcuoglu et~al.}(2006)\citenamefont{Gumrukcuoglu,
  Contaldi, and Peloso}}]{Gumrukcuoglu:2006xj}
\bibinfo{author}{\bibfnamefont{A.~E.} \bibnamefont{Gumrukcuoglu}},
  \bibinfo{author}{\bibfnamefont{C.~R.} \bibnamefont{Contaldi}},
  \bibnamefont{and} \bibinfo{author}{\bibfnamefont{M.}~\bibnamefont{Peloso}}
  (\bibinfo{year}{2006}), \eprint{astro-ph/0608405}.

\bibitem[{\citenamefont{Pullen and Kamionkowski}(2007)}]{Pullen:2007tu}
\bibinfo{author}{\bibfnamefont{A.~R.} \bibnamefont{Pullen}} \bibnamefont{and}
  \bibinfo{author}{\bibfnamefont{M.}~\bibnamefont{Kamionkowski}},
  \bibinfo{journal}{Phys. Rev.} \textbf{\bibinfo{volume}{D76}},
  \bibinfo{pages}{103529} (\bibinfo{year}{2007}), \eprint{0709.1144}.

\bibitem[{\citenamefont{Ando and Kamionkowski}(2007)}]{Ando:2007hc}
\bibinfo{author}{\bibfnamefont{S.}~\bibnamefont{Ando}} \bibnamefont{and}
  \bibinfo{author}{\bibfnamefont{M.}~\bibnamefont{Kamionkowski}}
  (\bibinfo{year}{2007}), \eprint{arXiv:0711.0779 [astro-ph]}.

\bibitem[{\citenamefont{Boehmer and Mota}(2007)}]{Boehmer:2007ut}
\bibinfo{author}{\bibfnamefont{C.~G.} \bibnamefont{Boehmer}} \bibnamefont{and}
  \bibinfo{author}{\bibfnamefont{D.~F.} \bibnamefont{Mota}}
  (\bibinfo{year}{2007}), \eprint{arXiv:0710.2003 [astro-ph]}.

\bibitem[{\citenamefont{Pitrou et~al.}(2008)\citenamefont{Pitrou, Pereira, and
  Uzan}}]{Pitrou:2008gk}
\bibinfo{author}{\bibfnamefont{C.}~\bibnamefont{Pitrou}},
  \bibinfo{author}{\bibfnamefont{T.~S.} \bibnamefont{Pereira}},
  \bibnamefont{and} \bibinfo{author}{\bibfnamefont{J.-P.} \bibnamefont{Uzan}},
  \bibinfo{journal}{JCAP} \textbf{\bibinfo{volume}{0804}}, \bibinfo{pages}{004}
  (\bibinfo{year}{2008}), \eprint{0801.3596}.

\bibitem[{\citenamefont{Dulaney et~al.}(2008)\citenamefont{Dulaney, Gresham,
  and Wise}}]{Dulaney:2008ph}
\bibinfo{author}{\bibfnamefont{T.~R.} \bibnamefont{Dulaney}},
  \bibinfo{author}{\bibfnamefont{M.~I.} \bibnamefont{Gresham}},
  \bibnamefont{and} \bibinfo{author}{\bibfnamefont{M.~B.} \bibnamefont{Wise}}
  (\bibinfo{year}{2008}), \eprint{arXiv:0801.2950 [astro-ph]}.

\bibitem[{\citenamefont{Carroll and Tam}(2008)}]{Carroll:2008pk}
\bibinfo{author}{\bibfnamefont{S.~M.} \bibnamefont{Carroll}} \bibnamefont{and}
  \bibinfo{author}{\bibfnamefont{H.}~\bibnamefont{Tam}} (\bibinfo{year}{2008}),
  \eprint{0802.0521}.

\bibitem[{\citenamefont{Hao and Li}(2003)}]{Hao:2003jm}
\bibinfo{author}{\bibfnamefont{J.-g.} \bibnamefont{Hao}} \bibnamefont{and}
  \bibinfo{author}{\bibfnamefont{X.-z.} \bibnamefont{Li}}
  (\bibinfo{year}{2003}), \eprint{hep-th/0303110}.

\bibitem[{\citenamefont{Mukhanov}(2005)}]{Mukhanov}
\bibinfo{author}{\bibfnamefont{V.}~\bibnamefont{Mukhanov}},
  \emph{\bibinfo{title}{Physical Foundations of Cosmology}}
  (\bibinfo{publisher}{Cambridge University Press}, \bibinfo{year}{2005}).

\bibitem[{\citenamefont{Ma and Bertschinger}(1995)}]{Ma:1995ey}
\bibinfo{author}{\bibfnamefont{C.-P.} \bibnamefont{Ma}} \bibnamefont{and}
  \bibinfo{author}{\bibfnamefont{E.}~\bibnamefont{Bertschinger}},
  \bibinfo{journal}{Astrophys. J.} \textbf{\bibinfo{volume}{455}},
  \bibinfo{pages}{7} (\bibinfo{year}{1995}), \eprint{astro-ph/9506072}.

\bibitem[{\citenamefont{Bertschinger}(2001)}]{Bertschinger:2001is}
\bibinfo{author}{\bibfnamefont{E.}~\bibnamefont{Bertschinger}}
  (\bibinfo{year}{2001}), \eprint{astro-ph/0101009}.

\bibitem[{\citenamefont{Misner et~al.}(1973)\citenamefont{Misner, Thorne, and
  J.Wheeler}}]{MTW:1973}
\bibinfo{author}{\bibfnamefont{C.}~\bibnamefont{Misner}},
  \bibinfo{author}{\bibfnamefont{K.}~\bibnamefont{Thorne}}, \bibnamefont{and}
  \bibinfo{author}{\bibnamefont{J.Wheeler}},
  \emph{\bibinfo{title}{Gravitation}} (\bibinfo{publisher}{W.H. Freeman and
  Company}, \bibinfo{year}{1973}).

\bibitem[{\citenamefont{Koivisto and Mota}(2007)}]{Koivisto:2007bp}
\bibinfo{author}{\bibfnamefont{T.}~\bibnamefont{Koivisto}} \bibnamefont{and}
  \bibinfo{author}{\bibfnamefont{D.~F.} \bibnamefont{Mota}}
  (\bibinfo{year}{2007}), \eprint{0707.0279}.

\bibitem[{\citenamefont{{Yao \emph{et.~al.}}}(2006)}]{PDBook}
\bibinfo{author}{\bibfnamefont{W.-M.} \bibnamefont{{Yao \emph{et.~al.}}}},
  \bibinfo{journal}{{Journal of Physics G}} \textbf{\bibinfo{volume}{33}},
  \bibinfo{pages}{1+} (\bibinfo{year}{2006}),
  \urlprefix\url{http://pdg.lbl.gov}.

\end{thebibliography}

\end{document}